\documentclass[12pt]{article}
\usepackage{amsmath}
\usepackage{amssymb}
\usepackage{amsfonts}

\begin{document}

\title{The deformation-stability fundamental length and deviations from c}
\date{}
\author{R. Vilela Mendes\thanks{%
CMAF, Instituto para a Investiga\c{c}\~{a}o Interdisciplinar, Av. Gama Pinto
2, 1649-003 Lisboa, Portugal, vilela@cii.fc.ul.pt,
http://label2.ist.utl.pt/vilela/} \thanks{%
IPFN - EURATOM/IST Association, Instituto Superior T\'{e}cnico, Av. Rovisco
Pais 1, 1049-001 Lisboa, Portugal}}
\maketitle

\begin{abstract}
The existence of a fundamental length (or fundamental time) has been
conjectured in many contexts. However, the "stability of physical theories
principle" seems to be the one that provides, through the tools of algebraic
deformation theory, an unambiguous derivation of \ the stable structures
that Nature might have chosen for its algebraic framework. It is well-known
that $1/c$ and $\hbar $ are the deformation parameters that stabilize the
Galilean and the Poisson algebra. When the stability principle is applied to
the Poincar\'{e}-Heisenberg algebra, two deformation parameters emerge which
define two length (or time) scales. In addition there are, for each of them,
a plus or minus sign possibility in the relevant commutators. One of the
deformation length scales, related to non-commutativity of momenta, is
probably related to the Planck length scale but the other might be much
larger. In this paper this is used as a working hypothesis to compute
deviations from $c$ in speed measurements of massless wave packets.
\end{abstract}


\section{Introduction}

The idea of modifying the algebra of the space-time components $x_{\mu }$ in
such a way that they become non-commuting operators has appeared many times
in the physical literature (\cite{Snyder} \cite{Yang} \cite{Kadyshevsky1} 
\cite{Kadyshevsky2} \cite{Banai}, \cite{Das} \cite{Atkinson} \cite{Gudder} 
\cite{Finkelstein} \cite{Dineykhan} \cite{Lee} \cite{Prugovecki} \cite%
{Blokhintsev} \cite{Schild} \cite{Veneziano} \cite{t Hooft} \cite{Jackson} 
\cite{Madore} \cite{Maggiore} \cite{Kempf}, etc.). The aim of most of these
proposals was to endow space-time with a discrete structure, to be able, for
example, to construct quantum fields free of ultraviolet divergences.
Sometimes a non-zero commutator is simply postulated, in some other
instances the motivation is the formulation of field theory in curved
spaces. String theories \cite{string} \cite{JHEP} and quantum relativity 
\cite{quantrelat} \cite{Dubois} have also provided hints concerning the
non-commutativity of space-time at a fundamental level.

A somewhat different point of view has been proposed in \cite{Vilela1} \cite%
{Vilela3}. There the space-time noncommutative structure is arrived at
through the application of the \textit{stability of physical theories
principle} (SPT). The rationale behind this principle is the fact that the
parameters entering in physical theories are never known with absolute
precision. Therefore, robust physical laws with a wide range of validity can
only be those that do not change in a qualitative manner under a small
change of parameters, that is, \textit{stable} (or \textit{rigid}) theories.
The stable-model point of view originated in the field of non-linear
dynamics, where it led to the notion of \textit{structural stability }\cite%
{Andronov} \cite{Smale}. Later on, Flato \cite{Flato} and Faddeev \cite%
{Faddeev} have shown that the same pattern occurs in the fundamental
theories of Nature, namely the transition from non-relativistic to
relativistic and from classical to quantum mechanics, may be interpreted as
the replacement of two unstable theories by two stable ones. The stabilizing
deformations lead, in the first case, from the Galilean to the Lorentz
algebra and, in the second one, from the algebra of commutative phase-space
to the Moyal-Vey algebra (or equivalently to the Heisenberg algebra). The
deformation parameters are $\frac{1}{c}$ (the inverse of the speed of light)
and $h$ (the Planck constant). Except for the isolated zero value, the
deformed algebras are all equivalent for non-zero values of $\frac{1}{c}$
and $h$. Hence, relativistic mechanics and quantum mechanics might have been
derived from the conditions for stability of two mathematical structures,
although the exact values of the deformation parameters cannot be fixed by
purely algebraic considerations. Instead, the deformation parameters are
fundamental constants to be obtained from experiment and, in this sense, not
only is deformation theory the theory of stable theories, it is also the
theory that identifies the fundamental constants.

The SPT principle is related to the idea that physical theories drift
towards simple algebras \cite{Segal} \cite{Finkelstein2} \cite{Finkelstein3}%
, because all simple algebras are stable, although not all stable algebras
are simple.

When the SPT principle is applied to the algebra of relativistic quantum
mechanics (the Poincar\'{e}-Heisenberg algebra)%
\begin{equation}
\begin{array}{rcl}
\lbrack M_{\mu \nu },M_{\rho \sigma }] & = & i(M_{\mu \sigma }\eta _{\nu
\rho }+M_{\nu \rho }\eta _{\mu \sigma }-M_{\nu \sigma }\eta _{\mu \rho
}-M_{\mu \rho }\eta _{\nu \sigma }) \\ 
\lbrack M_{\mu \nu },p_{\lambda }] & = & i(p_{\mu }\eta _{\nu \lambda
}-p_{\nu }\eta _{\mu \lambda }) \\ 
\lbrack M_{\mu \nu },x_{\lambda }] & = & i(x_{\mu }\eta _{\nu \lambda
}-x_{\nu }\eta _{\mu \lambda }) \\ 
\lbrack p_{\mu },p_{\nu }] & = & 0 \\ 
\lbrack x_{\mu },x_{\nu }] & = & 0 \\ 
\lbrack p_{\mu },x_{\nu }] & = & i\eta _{\mu \nu }\boldsymbol{1}%
\end{array}
\label{I.1}
\end{equation}%
$\eta _{\mu \nu }=(1,-1,-1,-1)$, $c=\hbar =1$, it leads \cite{Vilela1} to%
\begin{equation}
\begin{array}{rcl}
\lbrack M_{\mu \nu },M_{\rho \sigma }] & = & i(M_{\mu \sigma }\eta _{\nu
\rho }+M_{\nu \rho }\eta _{\mu \sigma }-M_{\nu \sigma }\eta _{\mu \rho
}-M_{\mu \rho }\eta _{\nu \sigma }) \\ 
\lbrack M_{\mu \nu },p_{\lambda }] & = & i(p_{\mu }\eta _{\nu \lambda
}-p_{\nu }\eta _{\mu \lambda }) \\ 
\lbrack M_{\mu \nu },x_{\lambda }] & = & i(x_{\mu }\eta _{\nu \lambda
}-x_{\nu }\eta _{\mu \lambda }) \\ 
\lbrack p_{\mu },p_{\nu }] & = & -i\frac{\epsilon ^{^{\prime }}}{R^{2}}%
M_{\mu \nu } \\ 
\lbrack x_{\mu },x_{\nu }] & = & -i\epsilon \ell ^{2}M_{\mu \nu } \\ 
\lbrack p_{\mu },x_{\nu }] & = & i\eta _{\mu \nu }\Im \\ 
\lbrack p_{\mu },\Im ] & = & -i\frac{\epsilon ^{^{\prime }}}{R^{2}}x_{\mu }
\\ 
\lbrack x_{\mu },\Im ] & = & i\epsilon \ell ^{2}p_{\mu } \\ 
\lbrack M_{\mu \nu },\Im ] & = & 0%
\end{array}
\label{I.2}
\end{equation}%
The stabilization of the Poincar\'{e}-Heisenberg algebra has been further
studied and extended in \cite{Chrysso} \cite{Ahluwalia1} \cite{Ahluwalia2}.
The essential message from (\ref{I.2}) or from the slightly more general
form obtained in \cite{Chrysso} is that from the unstable Poincar\'{e}%
-Heisenberg algebra $\{M_{\mu \nu },p_{\mu },x_{\nu }\}$ one obtains a
stable algebra with two deformation parameters $\ell $ and $\frac{1}{R}$. In
addition there are two undetermined signs $\epsilon $ and $\epsilon ^{\prime
}$and the central element of the Heisenberg algebra becomes a non-trivial
operator $\Im $. The existence of two continuous deformation parameters when
the algebra is stabilized is a novel feature of the deformation point of
view, which does not appear in other noncommutative space-time approaches.
These deformation parameters may define two different length scales. Of
course, once one of them is identified as a fundamental constant, the other
will be a pure number.

Being associated to the noncommutativity of the generators of space-time
translations, the parameter $\frac{1}{R}$ may be associated to space-time
curvature and therefore might not be relevant for considerations related to
the tangent space. It is, of course, very relevant for quantum gravity
studies \cite{Ahluwalia2}. Already in the past, some authors \cite{Faddeev},
have associated the noncommutativity of translations to gravitational
effects, the gravitation constant being the deformation parameter.
Presumably then $\frac{1}{R}$ might be associated to the Planck length
scale. However $\ell $, the other deformation parameter, defines a
completely independent length scale which might be much closer to laboratory
phenomena. This will be the working hypothesis to be explored in this paper.
Therefore when $\frac{1}{R}$ is assumed to be very small the deformed
algebra may be approximated by%
\begin{equation}
\begin{array}{rcl}
\lbrack M_{\mu \nu },M_{\rho \sigma }] & = & i(M_{\mu \sigma }\eta _{\nu
\rho }+M_{\nu \rho }\eta _{\mu \sigma }-M_{\nu \sigma }\eta _{\mu \rho
}-M_{\mu \rho }\eta _{\nu \sigma }) \\ 
\lbrack M_{\mu \nu },p_{\lambda }] & = & i(p_{\mu }\eta _{\nu \lambda
}-p_{\nu }\eta _{\mu \lambda }) \\ 
\lbrack M_{\mu \nu },x_{\lambda }] & = & i(x_{\mu }\eta _{\nu \lambda
}-x_{\nu }\eta _{\mu \lambda }) \\ 
\lbrack p_{\mu },p_{\nu }] & = & 0 \\ 
\lbrack x_{\mu },x_{\nu }] & = & -i\epsilon \ell ^{2}M_{\mu \nu } \\ 
\lbrack p_{\mu },x_{\nu }] & = & i\eta _{\mu \nu }\Im \\ 
\lbrack p_{\mu },\Im ] & = & 0 \\ 
\lbrack x_{\mu },\Im ] & = & i\epsilon \ell ^{2}p_{\mu } \\ 
\lbrack M_{\mu \nu },\Im ] & = & 0%
\end{array}
\label{I.3}
\end{equation}%
Notice that in addition to the space-time non-commutative structure, there
is also a new non-trivial operator $\Im $ which replaces the central element
of the Heisenberg algebra. In particular this operator corresponds to an
additional component in the most general connections compatible with (\ref%
{I.3}) \cite{Vilela3}.

For future reference this algebra will be denoted $\mathcal{R}_{\ell ,\infty
}$. Notice that in relation to the more general deformation obtained in \cite%
{Chrysso}, we are also considering $\alpha _{3}=0$ (or $\beta =0$ in \cite%
{Ahluwalia2}). The nature of the sign $\epsilon $ has physical consequences.
If $\epsilon =+1$ time will have a discrete spectrum, whereas if $\epsilon
=-1$ it is when one the space coordinates is diagonalized that discrete
spectrum is obtained. In this sense if $\epsilon =+1$, $\ell $ might be
called "the fundamental time" and "the fundamental length" if $\epsilon =-1$.

General (noncommutative) geometry properties of the algebra (\ref{I.3}) have
been studied before \cite{Vilela3} as well as some other consequences \cite%
{Vilela2} \cite{Carlen} \cite{Vilela4} \cite{Dzhunu} \cite{Goldin}. Here the
emphasis will be on the hypothesis that $\ell $ defines a length scale much
larger than Planck's and on its consequences for deviations from $c$ in
speed measurements of massless wave packets.

In the Appendix, some explicit representations of the space-time algebra are
collected, which are useful for the calculations.

\section{Measuring the speed of wave-packets}

In the noncommutative context, because the space and the time coordinates
cannot be simultaneously diagonalized, speed can only be defined in terms of
expectation values, for example%
\begin{equation}
v_{\psi }^{i}=\frac{1}{\left\langle \psi _{t},\psi _{t}\right\rangle }\frac{d%
}{dt}\left\langle \psi _{t},x^{i}\psi _{t}\right\rangle  \label{sw1}
\end{equation}%
Here, one considers a normalized state $\psi $ with a small dispersion of
momentum around a central value $p$. At time zero%
\begin{equation}
\psi _{0}=\int \left\vert k^{0}\overset{\longrightarrow }{k}\alpha
\right\rangle f_{p}\left( k\right) d^{3}k  \label{sw2}
\end{equation}%
where $k^{0}=\sqrt{\left\vert \overset{\longrightarrow }{k}\right\vert
^{2}+m^{2}}$, $\alpha $ standing for the quantum numbers associated to the
little group of $k$ and $f_{p}\left( k\right) $ is a normalized function
peaked at $k=p$.

To obtain $\psi _{t}$ one should apply to $\psi _{0}$ the time-shift
operator. However this is not $p^{0}$ because%
\begin{equation}
e^{-iap^{0}}te^{iap^{0}}=t+a\Im  \label{sw3}
\end{equation}%
follows from 
\begin{equation}
\left[ p^{0},t\right] =i\Im  \label{sw4}
\end{equation}%
whereas a time-shift generator $\Gamma $ should satisfy%
\begin{equation}
\left[ \Gamma ,t\right] =i\mathbf{1}  \label{sw5}
\end{equation}%
In order $O\left( \ell ^{4}\right) $ one has%
\begin{equation}
\Gamma =p^{0}\Im ^{-1}-\frac{\epsilon }{3}\ell ^{2}\left( p^{0}\right)
^{3}\Im ^{-3}  \label{sw6}
\end{equation}%
because%
\begin{equation}
\left[ \Gamma ,t\right] =i\left( 1-\ell ^{4}\left( p^{0}\right) ^{4}\Im
^{-4}\right)  \label{sw7}
\end{equation}%
To obtain this result, use was made of $\left[ t,\Im ^{-1}\right]
=-i\epsilon \ell ^{2}p^{0}\Im ^{-2}$, which follows from $\left[ t,\Im \Im
^{-1}\right] =0$.

Now use a basis where the set $\left( p^{\mu },\Im \right) $ is diagonalized
and define%
\begin{equation}
\overset{\thicksim }{p^{\mu }}=\frac{p^{\mu }}{\Im }  \label{sw8}
\end{equation}%
$\overset{\thicksim }{p^{\mu }}$ is the momentum in units of $\Im $.

Therefore, in the same $O\left( \ell ^{4}\right) $ order%
\begin{equation}
\psi _{t}=\int \exp \left( -it\left( \overset{\thicksim }{p^{0}}-\frac{%
\epsilon }{3}\ell ^{2}\left( \overset{\thicksim }{p^{0}}\right) ^{3}\right)
\right) \left\vert \overset{\thicksim }{k^{0}}\overset{\thicksim }{k^{i}}%
\alpha \right\rangle f_{p}\left( \overset{\thicksim }{k}\right) d^{3}\overset%
{\thicksim }{k}  \label{sw9}
\end{equation}%
To compute the expectation value of $x^{i}$ one notices that from (\ref{A.8})%
\begin{equation}
x^{\mu }=i\left( \epsilon \ell ^{2}p^{\mu }\frac{\partial }{\partial \Im }%
-\Im \frac{\partial }{\partial p_{\mu }}\right)   \label{sw10}
\end{equation}%
using $\frac{\partial }{\partial \Im }=-\frac{p^{\nu }}{\Im ^{2}}\frac{%
\partial }{\partial \overset{\thicksim }{p^{\nu }}}$ one obtains%
\begin{equation}
x^{\mu }=-i\left( \frac{\partial }{\partial \overset{\thicksim }{p_{\mu }}}%
+\epsilon \ell ^{2}\left\{ \overset{\thicksim }{p^{\mu }}\overset{\thicksim }%
{p^{\nu }}\frac{\partial }{\partial \overset{\thicksim }{p^{\nu }}}\right\}
_{S}\right)   \label{sw11}
\end{equation}%
$\left\{ {}\right\} _{S}$ meaning symmetrization of the operators.

Now the expectation value of this operator in the state $\psi _{t}$ is
computed and taking the time derivative one obtains for the wave packet
speed in order $\ell ^{2}$%
\begin{equation}
v_{\psi }=\frac{\overset{\thicksim }{p}}{\overset{\thicksim }{p^{0}}}\left(
1-\epsilon \ell ^{2}\left( p^{0}\right) ^{2}\right) -\epsilon \ell
^{2}\left( \overset{\thicksim }{p}p^{0}+\left( \overset{\thicksim }{p}%
\right) ^{2}\frac{\overset{\thicksim }{p}}{\overset{\thicksim }{p^{0}}}%
\right)  \label{sw12}
\end{equation}%
$\ell ^{2}$ being small, this deviation from $\frac{\overset{\thicksim }{p}}{%
\overset{\thicksim }{p^{0}}}$ may be difficult to detect for massive
particles given the uncertainty on the values of the masses. However, for
massless particles the deviation from $c\left( =1\right) $%
\begin{equation}
\Delta v_{\psi }=-3\epsilon \ell ^{2}\left( p^{0}\right) ^{2}  \label{sw13}
\end{equation}%
might already be possible to detect accurately with present experimental
means. Such deviation above or below the speed $c$ (depending on the sign of 
$\epsilon $) would not imply any modification of the relativistic
deformation constant $\left( \frac{1}{c}\right) $, nor a breakdown of
relativity.

The deviation formula (\ref{sw13}) will now be compared with the velocity
data of neutrino packets. The OPERA\ experiment \cite{OPERA} finds%
\begin{equation}
\frac{v-c}{c}=\left( 2.37\pm 0.32\right) \times 10^{-5}  \label{sw14}
\end{equation}%
for $p^{0}=17$ GeV neutrinos. Comparing with (\ref{sw13}) implies%
\begin{equation}
\ell =3.26%
\begin{array}{c}
+0.21 \\ 
-0.23%
\end{array}%
\times 10^{-18}cm  \label{sw15}
\end{equation}%
and $\epsilon =-1$. On the other hand the MINOS \cite{MINOS} result%
\begin{equation}
\frac{v-c}{c}=\left( 5.1\pm 2.9\right) \times 10^{-5}  \label{sw16}
\end{equation}%
using the peak energy at $3$ GeV would imply $\ell =2.7\times 10^{-17}cm$.
However this data has a long high energy tail, hence this result is not
reliable.

On the other hand the analysis of the SN1987A supernova \cite{Hirata} \cite%
{Bionta}, taking into account the fact that the outburst of visible light
begins later than the neutrino burst, when the cooler envelope is blown way,
led to the bound \cite{Longo}%
\begin{equation}
\left\vert \frac{v-c}{c}\right\vert \lesssim 2\times 10^{-9}  \label{sw17}
\end{equation}%
With the SN1987A neutrinos in the range $20-40$ MeV, one computes from (\ref%
{sw13}) with $p^{0}=40$ MeV and $\ell =3.2\times 10^{-18}$cm%
\begin{equation*}
\Delta v_{\psi }=1.26\times 10^{-10}
\end{equation*}%
compatible with the bound (\ref{sw17}). Incidentally, the shift for the
visible light is much smaller, namely $\Delta v_{\psi }=4.9\times 10^{-25}$
for $p^{0}=2.5$ eV.

In conclusion: the formula (\ref{sw13}) is consistent with all existing data
and the most probable value for $\ell $ is around $3\times 10^{-18}$ cm (or $%
10^{-28}$ s) and $\epsilon =-1$.

\section{Remarks and conclusions}

1) The most relevant point of the stability approach to noncommutative
space-time is the emergence of two deformation parameters, which might
define different length scales. This led to the conjecture that one of them
might be much larger than the Planck length and therefore already detectable
with contemporary experimental means.

2) The deviation from $c$ when measuring the velocity of wavepackets of
massless (or near massless) particles, does not implies any violation of
relativity nor does it imply a modification of the value of the deformation
parameter $\frac{1}{c}$. What it perhaps implies is that $c$ should not be
called the speed of light.

3) Other effects arising from the deformation-stability noncommutative
structure are explored in Ref. \cite{Vilela5}. Both the effects explore here
and in \cite{Vilela5} are rather conservative in the sense that they explore
well-known physical observables. A more speculative aspect of the
noncommutative structure concerns the physical relevance of the extra
derivation $\partial _{4}$ in the geometrical structure. This includes new
fields associated to gauge interactions which may lead to effective mass
terms for otherwise massless particles (see \cite{Vilela3} for more details)

\section{Appendix A: Representations of the deformed algebra and its
subalgebras}

For explicit calculations of the consequences of the non-commutative
space-time algebra (\ref{I.2}) (with $\epsilon ^{\prime }=0$) it is useful
to have at our disposal functional representations of this structure. Such
representations on the space of functions defined on the cone $C^{4}$ ($%
\epsilon =-1$) or $C^{3,1}$ ($\epsilon =+1$) have been described in \cite%
{Vilela3}. Here one collects a few other useful representations of the full
algebra and some subalgebras.

1 - As differential operators in a 5-dimensional commutative manifold $%
M_{5}=\{\xi _{\mu }\}$ with metric $\eta _{aa}=(1,-1,-1,-1,\epsilon )$%
\begin{equation}
\begin{array}{lll}
p_{\mu } & = & i\frac{\partial }{\partial \xi ^{\mu }} \\ 
\Im & = & 1+i\ell \frac{\partial }{\partial \xi ^{4}} \\ 
M_{\mu \nu } & = & i(\xi _{\mu }\frac{\partial }{\partial \xi ^{\nu }}-\xi
_{\nu }\frac{\partial }{\partial \xi ^{\mu }}) \\ 
x_{\mu } & = & \xi _{\mu }+i\ell (\xi _{\mu }\frac{\partial }{\partial \xi
^{4}}-\epsilon \xi ^{4}\frac{\partial }{\partial \xi ^{\mu }})%
\end{array}
\label{A.7}
\end{equation}

2 - Another global representation is obtained using the commuting set $%
\left( p^{\mu },\Im \right) $, namely%
\begin{equation}
\begin{array}{ccc}
x_{\mu } & = & i\left( \epsilon \ell ^{2}p_{\mu }\frac{\partial }{\partial
\Im }-\Im \frac{\partial }{\partial p^{\mu }}\right) \\ 
M_{\mu \nu } & = & i\left( p_{\mu }\frac{\partial }{\partial p^{\nu }}%
-p_{\nu }\frac{\partial }{\partial p^{\mu }}\right)%
\end{array}
\label{A.8}
\end{equation}

3 - Representations of subalgebras

Because of non-commutativity only one of the coordinates can be
diagonalized. Here, consider the restriction to one space dimension, namely
the algebra of $\left\{ p^{0},\Im ,p^{1},x^{0},x^{1}\right\} $.

For $\epsilon =+1$ define hyperbolic coordinates in the plane $\left(
p^{1},\Im \right) $ and polar coordinates in the plane $\left( p^{0},\Im
\right) $. Then, from it follows from (\ref{A.8})%
\begin{equation}
\begin{array}{lll}
p^{1} & = & \frac{r}{\ell }\sinh \mu \\ 
p^{0} & = & \frac{\gamma }{\ell }\sin \theta \\ 
\Im & = & r\cosh \mu =\gamma \cos \theta \\ 
x^{1} & = & i\ell \frac{\partial }{\partial \mu } \\ 
x^{0} & = & -i\ell \frac{\partial }{\partial \theta }%
\end{array}
\label{A.9}
\end{equation}%
For $\epsilon =-1$ with polar coordinates in the plane $\left( p^{1},\Im
\right) $ and hyperbolic coordinates in the plane $\left( p^{0},\Im \right) $%
,%
\begin{equation}
\begin{array}{lll}
p^{1} & = & \frac{r}{\ell }\sin \theta \\ 
p^{0} & = & \frac{\gamma }{\ell }\sinh \mu \\ 
\Im & = & \gamma \cosh \mu =r\cos \theta \\ 
x^{1} & = & i\ell \frac{\partial }{\partial \theta } \\ 
x^{0} & = & -i\ell \frac{\partial }{\partial \mu }%
\end{array}
\label{A.10}
\end{equation}

\end{document}